\documentclass[journal]{IEEEtran}
\usepackage{blindtext}
\usepackage{graphicx}
\usepackage{amssymb,amsmath,amsthm}
\let\labelindent\relax
\usepackage{enumitem}
\usepackage{placeins}
\usepackage{hhline,makecell}
\usepackage{algorithm, algcompatible}
\usepackage{algpseudocode}
\usepackage{tikz}
\usepackage[T1]{fontenc}
\usepackage[utf8x]{inputenc}
\usepackage{relsize}

\tikzset{fontscale/.style = {font=\relsize{#1}}
    }
\usetikzlibrary{automata, positioning}
\usepackage[algo2e]{algorithm2e} 
\theoremstyle{plain}
\newtheorem{thm}{Lemma}
\newtheorem{thm2}[thm]{Theorem}

\ifCLASSINFOpdf

\else

\fi

\begin{document}

\title{Meeting of Mobile Nodes Based on RSS Measurements in Wireless Ad Hoc Networks\vspace{0.3cm}}

\author{
        \IEEEauthorblockN{Noyan Evirgen\IEEEauthorrefmark{1}, Alper Köse\IEEEauthorrefmark{2}\IEEEauthorrefmark{3}}
        \IEEEauthorblockA{
                \small\\
                 \vspace{0.12cm}
                \IEEEauthorrefmark{1}Department of Information Technology and Electrical Engineering, ETH Zürich\\
                \vspace{-0.1cm}
                \IEEEauthorrefmark{2}Department of Electrical Engineering, École Polytechnique Fédérale de Lausanne\\ 
                \vspace{0.1cm}
                \IEEEauthorrefmark{3}Research Laboratory of Electronics, Massachusetts Institute of Technology\\
                \vspace{0.11cm}
                nevirgen@student.ethz.ch, akose@mit.edu
                \vspace{-0.2cm}
        }
    
}

\maketitle

\begin{abstract}
In this work, we address a completely novel case which is the meeting problem of mobile nodes (players) in a wireless ad hoc network. We assume that the only information players have is the received signal strength (RSS) measurements of other players and there is no centralized communication. Each node has a different frequency band and they try to meet with each other in a decentralized manner. We consider the case where players are allowed to move in four orthogonal directions in a grid. Then, we use multi-armed bandit approach to give rewards in the end of every move that players do using RSS measurements. Taking into account that we have an adversarial setting, we propose different algorithms to solve the meeting problem. We start with considering two players and then extend the work to multiplayer by considering multiplayer problem as multiple two player problems. Throughout the paper, we construct mathematical bounds on meeting times and related concepts and evaluate the simulation results in the end.

\end{abstract}


\IEEEpeerreviewmaketitle

\section{Introduction}
\label{sec:intro}

We consider a meeting problem in mobile ad hoc networks. In our setting, the goal of the nodes is to meet with each other using only received signal strength (RSS) measurements. We assume that there is no communication among the players and they aim to find each other in a decentralized manner. This problem is challenging in the sense that there is no centralized system that decides on the moves. Importantly, it is not possible to localize other players with just using the signal strength received from them since there needs to be at least three nodes communicating to solve the triangulation problem. However, we have a mobile ad hoc network and as the players move it becomes possible for them to have an idea of other players' positions. Localization problem in ad hoc networks have been studied thoroughly in \cite{savvides2001dynamic}, \cite{niculescu2003ad}, \cite{biswas2004semidefinite}, \cite{niculescu2003dv}, \cite{gezici2008survey}. For instance Zickler and Veloso \cite{zickler2010rss} use RSS for localization and tethering of the robots with the addition of odometry data. In their work, a robot follows another robot via localization. Even though initially it is not possible for the first robot to localize the second robot, as the robots move, the first robot can find the other one's position using a sequential Monte Carlo method. Another example is the LANdroids program where the goal is to enhance the tactical communication in urban environments \cite{mcclure2009darpa}. LANdroids move in order to establish communication among soldiers and this setup requires an autonomous system. These droids find soldiers for better throughput across the network. 

In this work, we focus on a different aspect of mobile ad hoc networks where the goal is the meeting of the players which have use cases in robotics, military, maintaining communication in mobile ad hoc networks. In particular, consider a group of soldiers scattered to a forest due to an enemy attack in a battle field. How can they merge again? We address this problem given that every soldier is equipped with a radio in order to only measure signal strengths of other radios. Note that, we assume that there is no meaningful communication between radios but only the signal strengths of other radios can be monitored. In short, our contribution is to propose a feasible strategy aiming to bring all mobile nodes together in a wireless ad hoc network as earliest as possible using only RSS measurements in a decentralized manner.

In a mobile ad hoc network, it is beneficial to have a robust meeting algorithm for relay nodes so that connections among nodes are maintained. This becomes especially challenging in low SNR environments with multipath effects. Under this setting we investigate the meeting of the nodes in mobile ad hoc networks based on RSS measurements. We further limit our setting with the assumption of having no communication between nodes since it is not always possible to have high SNR to maintain a connection. Each and every player chooses one of the four directions, and makes their moves every turn, which lead to different RSS measurements. These measurements are used to define rewards. Each node uses multi-armed bandit approach for making decisions on how to move. Multi-armed bandits can be used in various settings and proved to be a successful approach in modelling problems. Some examples of multi-armed bandit problems can be found in \cite{bagheri2015restless}, \cite{liu2010distributed}, \cite{radlinski2008learning}. Also, Pini et al. \cite{pini2012multi} propose a multi-armed bandit formulation of task completion problem for robots and it is similar to our work in the sense that we model the actions of people instead of robots. On the other hand, it can be observed that the possible rewards for each player may change in time. For instance, relative positions of players will possibly change and this will alter the reward distribution on arms from the perspective of a player. Therefore, our problem should be treated as an adversarial case \cite{auer2002nonstochastic}. In the solution of our problem, we firstly consider the case of two players. Then, we extend it to the multiplayer case by modelling the multiplayer problem as multiple two player problems. 

The paper is organized as follows. We explain the change of signals along their paths and give the models we use in Section \ref{sec:network}. Formulation of the problem from a multi-armed bandit perspective and an exemplification for two players is shown in Section \ref{sec:algos}. Performance evaluation of different algorithms under the assumptions of our model can be seen in Section \ref{algolarrr}. We comment on the simulation results in Section \ref{simres} and conclude the paper in Section \ref{concon}.

\section{Network Model}
\label{sec:network}

A signal while traveling from one node to another decays due to multipath fading, shadowing and path loss \cite{gezici2008survey}. When the RSS is averaged over a sufficient amount of time the effects of multipath fading can be mitigated. The path loss can be modeled as a log-distance loss model which is,
\begin{equation}
\bar{P}_d = P_i-10 n_p \log\left(\dfrac{d}{d_0}\right)
\end{equation}
where $n_p$ is the path loss exponent, $\bar{P}_d$ is the average received power in dB at a distance $d$ due to path loss and $P_i$ is the received power in dB at a reference distance $d_0$. In practice the effect of shadowing can not be mitigated with averaging which can be modeled by a Gaussian random variable with a variance of $\sigma$~\cite{gezici2008survey}. Thus, the received power at a distance $d$ is,

\begin{equation}
P_d = \mathcal{N}(P_i-10 n_p \log\left(\dfrac{d}{d_0}\right),\sigma^{2})
\end{equation}

Moreover, shadowing is a first-order auto-regressive process~\cite{goldsmith2005wireless}. The covariance between shadow fading of two points separated by distance $\delta$ is,
\begin{equation}
A(\delta) = \sigma^{2}e^{-\delta/X_c}
\end{equation}

where $A(\delta)$ is the signal auto-covariance and $X_c$ is decorrelation distance which is on the order of size of the blocking objects. For outdoor systems it typically ranges between 50-100 meters \cite{goldsmith2005wireless}. When the players move around their RSS decay due to shadowing between adjacent turns are correlated. Thus, correlation of shadowing is crucial in our setting.

\section{Problem Formulation}
\label{sec:algos}

We model the problem as a multi-armed bandit (MAB) problem. In each turn players have four different options; \textbf{up}, \textbf{right}, \textbf{down} and \textbf{left}. These four options can be considered as arms where each player chooses an arm and receives a reward. Players do not know each other's decisions directly but they can estimate it by observing the change in the received signal strength.  

In order to assess the quality of the arms, we define the reward metric. These four arms have different reward distributions for each turn based on the position of the players. The reward at each turn can be defined as:
\begin{equation}
R_{k}^l(t) = S_{k}^l(t)-S_{k}^l(t-1)
\end{equation}
for $t = \{2,3,..\}$, where $S_{k}^l(t)$ is the received signal strength at the $k$th player which is sent by the $l$th player at time $t$.

Note that each turn players are allowed to walk $\Delta$ in one of the four different directions. Let us assume there are two players and let us denote $R_1^2(t)$ and $S_1^2(t)$ with $R(t)$ and $S(t)$ respectively. Potential rewards between two players can be considered in sixteen different cases. Assume that $\{x_t,y_t\}$ is the position of the second player relative to the first player at turn $t$. Then the received signal strength can be shown as:

\begin{equation}
\mathbb {P}(S(t)) = \mathcal{N}(P_i-10 n_p \log\left(\dfrac{\sqrt{x_t^2+y_t^2}}{d_0}\right),\sigma_t^{2})
\end{equation}

The RSS between adjacent turns can be calculated using the above equation as:

\begin{equation}
\begin{split}
\mathbb {P}(R(t+1)) = &\mathcal{N}(P_i-10 n_p \log\left(\dfrac{\sqrt{x_{t+1}^2+y_{t+1}^2}}{d_0}\right),\sigma_{t+1}^{2}) \\ - &\mathcal{N}(P_i-10 n_p \log\left(\dfrac{\sqrt{x_t^2+y_t^2}}{d_0}\right),\sigma_t^{2}) \\ = \mathcal{N}(-5 n_p \log&\left(\dfrac{x_{t+1}^2+y_{t+1}^2}{x_t^2+y_t^2}\right),\sigma_{t+1}^{2}+\sigma_{t}^{2}-2\sigma_{t(t+1)}) \\
= \mathcal{N}(-5 n_p \log&\left(\dfrac{x_{t+1}^2+y_{t+1}^2}{x_t^2+y_t^2}\right),2\sigma^{2}-2\sigma^2e^{-\delta/X_c})
\end{split}
\end{equation}
where $\sigma_{t(t+1)}$ is the covariance between the Gaussian random variables which represents shadowing at turn $t$ and $t+1$. Since both of the players move $\Delta$ in each turn, the covariance can be modeled using Wang \textit{et al.} model \cite{wang2006simulating}. Therefore, $\delta = 2\Delta$. The assumption is that the problem we have is equivalent to one player moving 2$\Delta$ instead of two players both moving $\Delta$ as far as shadowing correlation is concerned. We also assume that $\sigma = \sigma_t$ for any turn $t$. 

\section{Algorithms}
\label{algolarrr}
Four different algorithms are proposed for the problem of meeting in wireless networks. Before introducing the algorithms, the log distance path loss model is analyzed in our setting with the following lemmas. The proofs of the lemmas are at Appendix A. 

\begin{thm}
Assuming that the second player is moving randomly and $d_t \gg 2\Delta$ where $d_t = \sqrt{x_t^2+y_t^2}$, the expected RSS change between adjacent turns depends only on $x_t$ if the first player chooses to move along the x-axis or on $y_t$ if the first player chooses to move along the y-axis.
\end{thm}

The assumption of second player moving randomly is a fair assumption in our setting. This is due to the fact that the expected RSS change between adjacent turns is close to zero which is shown in the proof. Thus, players are moving almost randomly to find each other. Second assumption is always correct since players are trying to find each other in a big area with small steps.

\begin{thm}
Assuming players return to the previous position after each turn, probability of players finding the positive arms +x,+y after N turns is in the order of $\sqrt{N}$.
\end{thm}

\begin{figure}[!t]
  \centering
    \includegraphics[width=0.45\textwidth]{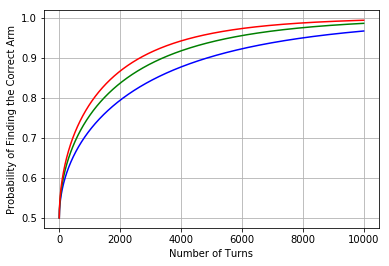}    
    \caption{Probability of finding the correct arm vs number of plays for $\Delta = 0.1$ meters, $\sigma = 4$, $n_p = 4$ and $X_c = 75$ meters. Blue line corresponds to $x_t = 40$ and $y_t = 70$. Green line corresponds to $x_t = 100$ and $y_t = 60$. Red line corresponds to $x_t = 60$ and $y_t = 60$.} 
    \label{fig:3line}
\end{figure}
Path loss model that is used in this paper already assumes RSS is averaged over sufficient amount of turns to get rid of multipath fading. Therefore, further averaging is costly in terms of time. However, assuming it is possible, from Lemma 2 it can be seen that further averaging is an inefficient solution to meeting. Using Equation \ref{eq:qfunction}, Figure \ref{fig:3line} shows the required number of plays for different certainties. As it can be seen, it is not possible to find the positive arms with reasonable amount of initial random plays. Moreover, assuming a correct arm is found, it is not clear how many steps a player needs to take in that direction. Players also do not know the model parameters and they are not synchronized to walk randomly until they both know the correct arm which makes this approach unfeasible.  

\subsection{Greedy Algorithm}

Because of all the mentioned reasons, the players have to move with uncertainty and still need to find each other. Therefore we introduce greedy algorithm where we employ Lemma 1 to decide on every other play. From Equation \ref{eq:rdist} and \ref{eq:rdist2}, it can be seen that if an arm has positive expected reward, its reverse arm has a negative expected reward. Greedy algorithm leverages this fact and pulls the reverse arm next turn if the current reward is negative. If the reward is positive, it pulls the same arm next turn. 

\begin{table}[!tb]
{\LinesNumberedHidden
    \begin{algorithm}[H]
        \SetKwInOut{Input}{Input}
        \SetKwInOut{Output}{Output}
        \SetAlgorithmName{Algorithm}{}
        \\
        $t$ is the turn number.\\
        $\mathcal{I}(t)$ is the chosen arm in turn $t$.\\
        $\mathcal{I}'(t)$ is the arm in the reverse direction of the chosen arm in turn $t$.\\
        Initialize:\\
        \If {t = 1} {
            Pull a random arm and denote it with $\mathcal{I}_1$.
        }
        \For {t = 2,3,...}{
            \If{t \textbf{mod} 2 = 0}{
                Pull a random arm and denote it with $\mathcal{I}(t)$.\\
                Observe the reward and denote the reward with $R(t)$.\\
                \If{$R(t) > 0$}{
                    $\mathcal{I}(t+1) = \mathcal{I}(t)$.
                }
                \Else{
                    $\mathcal{I}(t+1) = \mathcal{I}'(t)$.
                }
            }
            \If{t \textbf{mod} 2 = 1}{
                Pull arm according to $\mathcal{I}(t)$, which is denoted in the previous turn.
            }
        } 
\caption{Greedy Algorithm}
\label{algo:greedy}
\end{algorithm}}
\caption{Greedy Algorithm}
\end{table}

\begin{thm2} \label{greedy_turn}
Expected number of turns for meeting of the Greedy algorithm can be upper bounded by $\dfrac{(x_t^2+y_t^2)\sqrt{2\pi\sigma^2(1-e^{-2\Delta/X_c})}}{5 n_p \Delta^2}$ turns. 
\end{thm2}

where ($x_t$, $y_t$) is the initial relative position of the second player with respect to the first player. The Greedy algorithm is sub-optimal. Considering we have two players and the L1-distance between their positions is $x_t+y_t$, the total distance traversed by the players collectively, denoted as $D$, is at least $x_t+y_t$. The expected value of this total traversed distance, denoted by $E[D]$ can be calculated as follows,

\begin{equation}
    E[D] = E[T] \cdot 2\Delta
    \label{eq:traversed}
\end{equation}
where the upper-bound of expected number of turns $E[T]$ is given in Theorem \ref{greedy_turn} and $2\Delta$ represent total distance traversed collectively by 2 players.

As $\Delta \rightarrow 0$, $E[D]$ diverges to $+\infty$. For $\Delta > 0$, $E[D]$ is finite. Taking the first derivative of $E[D]$, we need
\begin{equation} \label{delta_der}
    \log \left(1+\frac{2\Delta}{X_c}\right) = \frac{2\Delta}{X_c}
\end{equation}
for a possible local minimum to exist for $\Delta > 0$. However, for $x > 0$, $\log(1+x) < x$. Hence, no local minimum exists and $E[D]$ is a strictly decreasing function of $\Delta \geq 0$.

Consequently, we need to choose $\Delta$ as high as possible. This may not seem intuitive. The reason can be explained via the fact that greedy algorithm uses information from the last two turns only and when we choose $\Delta$ very small the change in our space-dependent noise is also very small. Consequently, we are more prone to make incorrect movement decisions.

Unfortunately, we cannot choose $\Delta$ arbitrarily large since the derivation of Theorem \ref{greedy_turn} also required the assumption that $\Delta \ll d$. Therefore, for the occasions we cannot choose $\Delta \in \Theta(d)$, we cannot achieve $E[D] \in O(d)$ much less the optimum value, $E[D] = x+y \leq \sqrt{2} d$. $\Theta$ and $O$ are the big-Omega and big-O notations respectively.

As an example for the same conditions from Figure~\ref{fig:3line}, the upper bounds $E[T]$ are $16816$, $35184$ and $18627$ turns for $x_t=40,y_t=70$, $x_t=100,y_t=60$ and $x_t=60,y_t=60$ respectively. The experiments give $2370$, $3680$ and $2551$ turns respectively when averaged over 1000 trials with the same conditions. It can be seen that the values are in the same order of magnitude with the necessary amount of turns for finding the positive arms from Figure~\ref{fig:3line}. Considering finding the positive arms is not enough for meeting, meeting without finding the best arm is a better strategy. That is the intuition behind the greedy algorithm where players only use the information that is available from the last turn. 

\subsection{R-Exp3 Algorithm}

By relaxing the Greedy algorithm and allowing it to depend on finite amount of previous turns we get the R-Exp3 algorithm, in which the arm is chosen based on a sampling of exponential losses. R-Exp3 is a modified version of the original Exp3 algorithm \cite{auer2002nonstochastic}. As in the Greedy algorithm, Equation \ref{eq:rdist} and \ref{eq:rdist2} are used to modify Exp3. Since negative moves have negative means whereas positive moves have positive means, the sampling in the Exp3 can be modified such that arms in the opposite directions are coupled. The coupled arms which have bigger difference between their weights are sampled with higher chance since each couple has one positive and one negative arm. After deciding on the couple, in other words the axis, the sampling between the positive and negative arm is done as it is done in the Exp3 algorithm.
\begin{table}[!t]
{\LinesNumberedHidden
    \begin{algorithm}[H]
        \SetKwInOut{Input}{Input}
        \SetKwInOut{Output}{Output}
        \SetAlgorithmName{Algorithm}{}
        
        $t$ is the turn number.\\
        $i = 1,2,3,4$ corresponds to up, down, right, left arms respectively.\\
        Initialize:\\
        Select a parameter $a \in (0,1)$.\\
        $w_i(1) = 0.25$ $\forall$ $i \in \{1,2,3,4\}$.\\ 
        \For{t = 1,2,3,..}{
            $p_i(t) = (1-a)(\dfrac{w_i(t)}{\sum_{j=1}^4w_j(t)})+\dfrac{a}{4}$.\\
            \If{t \textbf{mod} 2 = 0}{
                Weights of the axis-es using the probabilities: \\ $h_1(t) = \dfrac{|p_1(t)-p_2(t)|+1/2}{|p_1(t)-p_2(t)|+|p_3(t)-p_4(t)|+1}$.\\ $h_2(t) = \dfrac{|p_3(t)-p_4(t)|+1/2}{|p_1(t)-p_2(t)|+|p_3(t)-p_4(t)|+1}$.\\ 
                Sample from the random variable $X_1(t)$ with probability mass function(pmf): $P(X_1(t) = i) = h_i(t)$.\\
                Denote the output with $L_1(t)$.\\
                Calculate coupled weights:\\
                \If{$L_1(t) == 1$}{
                    $k_1(t) = \dfrac{p_1(t)}{p_1(t)+p_2(t)}$, $k_2(t) = \dfrac{p_2(t)}{p_1(t)+p_2(t)}$.
                }
                \If{$L_1(t) == 2$}{
                    $k_1(t) = \dfrac{p_3(t)}{p_3(t)+p_4(t)}$, $k_2(t) = \dfrac{p_4(t)}{p_3(t)+p_4(t)}$.
                }
                Sample from $X_2(t)$ with pmf: $P(X_2(t) = i) = k_i(t)$.\\
                Denote the output $L_2(t)$.\\
                Choose arm index: $\mathcal{I}(t) = (L_1(t))^2 - L_1(t) + L_2(t)$.\\
                Observe the reward of $\mathcal{I}(t)$ and denote it with $R_1(t)$.\\
                \If{$R_1(t) > 0$}{
                    $\mathcal{I}(t+1) = \mathcal{I}(t)$.
                }
                \Else{
                    $\mathcal{I}(t+1) = \mathcal{I}'(t)$.
                }
            }
            \If{t \textbf{mod} 2 = 1}{
                Pull arm according to $\mathcal{I}(t)$. Observe reward R(t).
            }
            \If{$R(t) > 0$}{
                $R_2(t) = 1/p_{\mathcal{I}(t)}(t)$.
            }
            \Else{
                 $R_2(t) = 0$.
            }
            Update the distribution:\\
            $w_{\mathcal{I}(t)}(t+1) = w_{\mathcal{I}(t)}(t)\exp(aR_2(t)/4)$
        }

\caption{R-Exp3 Algorithm}
\end{algorithm}}
\caption{R-Exp3 Algorithm}
\label{algo:exp3}
\end{table}

The Exp3 algorithm has the advantage of decaying weights of the previous turns which makes it a suitable method for adversarial bandit problems. R-Exp3 is designed on top of the same principle and therefore has the same characteristic. Greedy algorithm is similar in the sense that we only use the rewards for the next turn and discard it afterwards, which makes it suitable for adversarial bandit problems. 

\subsection{R-Thompson Sampling Algorithm}

We use binary rewards in both Greedy and R-Exp3 Algorithm. The reason is, from Figure \ref{fig:dashed_area}, it can be seen that the reward distribution's mean is close to zero with high variance. Thus, using the actual rewards do not give additional information compared to binary. In addition previous two algorithms, we introduce a different algorithm. Since we have binary rewards, we introduce a Thompson sampling based algorithm called R-Thompson Sampling Algorithm \cite{agrawal2012analysis}.

\begin{table}[!t]
{\LinesNumberedHidden
    \begin{algorithm}[H]
        \SetKwInOut{Input}{Input}
        \SetKwInOut{Output}{Output}
        \SetAlgorithmName{Algorithm}{}
        
        $t$ is the turn number.\\
        $\epsilon$ is the learning rate.\\
        $i = 1,2,3,4$ corresponds to up, down, right, left arms respectively.\\
        Initialize:\\
        Set $S_i(1) = 0$, $F_i(1) = 0$ $\forall$ $i \in \{1,2,3,4\}$.\\
        \For{t = 1,2,3,...}{
            \If{t \textbf{mod} 2 = 0}{
                Sample $\theta_{i}(t)$ for each arm from the $Beta(S_{i}(t) +1,F_{i}(t) +1)$ distribution.\\
                Calculate $L_1(t) = argmax\{|\theta_1(t)-\theta_2(t)|,|\theta_3(t)-\theta_4(t)|\}$.\\
                Find the chosen arm index with:\\
                \If{$L_1(t) == 1$}{
                    $\mathcal{I}(t) = argmax\{\theta_1(t),\theta_2(t)\}$.
                }
                \If{$L_1(t) == 2$}{
                    $\mathcal{I}(t) = 2+argmax\{\theta_3(t),\theta_4(t)\}$.   
                }
                Observe the reward of $\mathcal{I}(t)$ and denote it with $R(t)$.
            }
            \If{t \textbf{mod} 2 = 1}{
                Pull arm according to $\mathcal{I}(t)$. Observe reward R(t).
            }
            Update the distribution:\\
            \If{$R(t) > 0$}{
                $S_{\mathcal{I}(t)}(t+1)$ = $S_{\mathcal{I}(t)}(t) + \epsilon$.
            }
            \Else{
                $F_{\mathcal{I}(t)}(t+1)$ = $F_{\mathcal{I}(t)}(t) + \epsilon$.
            }
        }
        
\caption{R-Thompson Sampling Algorithm}
\end{algorithm}}
\caption{R-Thompson Sampling Algorithm}
\label{algo:Thompson Sampling Algorithm}
\end{table}

The upper bound introduced for Greedy algorithm is also valid for R-Exp3 and R-Thompson algorithms, considering Greedy algorithm is a special case of them where the moves only depend on the last turn. Instead of the random turn of the Greedy algorithm, we have traditional bandit algorithms applied for exploration. 

\subsection{R-Optimal Algorithm}

Even though the traditional adversarial bandit algorithms work in our setting and their estimated meeting time can be bounded with O$\left(\dfrac{d^2\sigma(1-e^{-2\Delta/X_c})}{n_p\Delta^2}\right)$ turns and give reasonable performance, under mild assumptions the expected meeting time can be bounded by O($K\dfrac{d\sigma(1-e^{-2\Delta/X_c})}{n_p\Delta}+\dfrac{d}{\Delta}$) with an improved version of Greedy algorithm which is called R-Optimal algorithm. In R-Optimal algorithm, we use the rewards of each arm to calculate the reward means. Using these means, we decide on the arm for the exploration turn instead of randomly choosing it as in the Greedy algorithm. From Lemma 1 it follows that x-axis and y-axis are independent from each other, therefore before the turn starts we randomly choose an axis. Similar to previous algorithms, every other turn, if the reward is positive we choose the same arm again and if the reward is negative we choose the arm in the reverse direction. The algorithm is asymptotically optimal in the sense that once it 'converges' to the positive arms, it stays in that state with high probability. This is shown in the proof. 

\begin{thm2}\label{opt_turn}
Expected number of turns for meeting of the R-Optimal algorithm can be upper bounded with $\dfrac{2zd\sqrt{2\pi\sigma^2(1-e^{-2\Delta/X_c})}}{5n_p\Delta} + \dfrac{d}{\Delta}$ turns. 
\end{thm2}

where $d$ is $\sqrt{x_t^2+y_t^2}$ and $z$ is a constant. For $z>10$, the upper bound holds with high probability. Therefore, expected meeting time can be shown as O($K\dfrac{d\sigma(1-e^{-2\Delta/X_c})}{n_p\Delta}+\dfrac{d}{\Delta}$) where $K$ is a constant depending on $z$. This algorithm significantly outperforms the Greedy algorithm for the case where the initial distance is high, since R-Optimal has O($K_1d$) expected meeting time compared to O($K_2d^2$) of Greedy algorithm where $K_1$ and $K_2$ are constants. For the case of short distances, Greedy algorithm is expected to work better because of $K$ and the additional term, which is also validated by simulations. For the same conditions from Figure~\ref{fig:3line}, the upper bounds are $1223$, $1770$ and $1287$ turns for $x_t=40,y_t=70$, $x_t=100,y_t=60$ and $x_t=60,y_t=60$ respectively. The experiments give $882$, $1325$ and $936$ turns respectively when averaged over 100 trials with the same conditions.

The expected value of the total traversed distance, denoted by $E[D]$ can be calculated as in Equation \ref{eq:traversed}, where the upper-bound of expected number of turns $E[T]$ is given in Theorem \ref{opt_turn} and $2\Delta$ represent total distance traversed collectively by 2 players.It can be seen that, as $\Delta \rightarrow 0$, $E[D]$ converges to $2d$. The difference with respect to the Greedy algorithm is that since we are using all the information available in each turn, by decreasing $\Delta$, the performance of the algorithm improves. Therefore, for R-Optimal algorithm to work well, players need to take small steps.

\begin{table}[!t]
{\LinesNumberedHidden
    \begin{algorithm}[H]
        \SetKwInOut{Input}{Input}
        \SetKwInOut{Output}{Output}
        \SetAlgorithmName{Algorithm}{}
        
        $t$ is the turn number.\\
        $i = 1,2,3,4$ corresponds to up, down, right, left arms respectively.\\
        $\mathcal{I}(t)$ is the chosen arm in turn $t$.\\
        $\mathcal{I}'(t)$ is the arm in the reverse direction of the chosen arm in turn $t$.\\
        Initialize:\\
        Set $p_i(1) = 0$, $c_i(t) = 1$ $\forall$ $i \in \{1,2,3,4\}$.\\
        \For{t=1,2,3,...}{
            \If{t \textbf{mod} 2 = 0}{
                Calculate $k_i(t) = \dfrac{p_i(t)}{c_i(t)}$.\\
                Random a value between 0 and 1, denote it with $\alpha$.\\
                \If{$\alpha > 0.5$}{
                    $\mathcal{I}(t) = argmax\{k_1(t),k_2(t)\}$}
                \Else{
                    $\mathcal{I}(t) = argmax\{k_3(t),k_4(t)\}+2$
                }
                where $\mathcal{I}(t)$ denotes the chosen arm for turn $t$.\\
                Observe the reward of $\mathcal{I}(t)$ and denote it with $R(t)$.\\
                 \If{$R(t) > 0$}{
                    $\mathcal{I}(t+1) = \mathcal{I}(t)$.
                }
                \Else{
                    $\mathcal{I}(t+1) = \mathcal{I}'(t)$.
                }
            }
            \If{t \textbf{mod} 2 = 1}{
                Pull arm according to $\mathcal{I}(t)$, which is denoted in the previous turn.
            }
            \If{$R(t) > 0$}{
                $p_{\mathcal{I}(t)}(t+1) = p_{\mathcal{I}(t)} + 1$}
            $c_{\mathcal{I}(t)}(t+1) = c_{\mathcal{I}(t)} + 1$
        }
        
\caption{R-Optimal Algorithm}
\end{algorithm}}
\caption{R-Optimal Algorithm}
\label{algo:Thompson Sampling Algorithm}
\end{table}

\subsection{Extension to Multiplayer}

Throughout the paper algorithms are considered for two players. The extension of the algorithms to multiplayer setting is done by considering the problem as a multiple two player problems. Since there is no communication between players, they need to choose one of the other participating players and try to meet with them.  We assume that when two players meet, they start to move together to meet with other players. Considering the path loss model used is obtained by averaging RSS, it is assumed that further averaging is not possible. Under this assumption the difference between the RSS coming to the first player, from the second and the third players can be calculated as:

\begin{equation}
\begin{split}
\mathbb {P}(S_{1}^2(t)-S_{1}^3(t)) &= \mathcal{N}(P_i-10 n_p \log\left(\dfrac{d_{1,2}}{d_0}\right),\sigma_2^{2}) \\ &- \mathcal{N}(P_i-10 n_p \log\left(\dfrac{d_{1,3}}{d_0}\right),\sigma_3^{2}) \\ = \mathcal{N}(-10 n_p \log&\left(\dfrac{d_{1,2}}{d_{1,3}}\right),\sigma_{2}^{2}+\sigma_{3}^{2}-2\sigma_{(2)(3)}) \\
= \mathcal{N}(-10 n_p \log&\left(\dfrac{d_2}{d_3}\right),2\sigma^{2}-2\sigma^2e^{-d_{2,3}/X_c})
\end{split}
\end{equation}
where $d_{i,j}$ is the distance between the $i$th and $j$th player. $\sigma_2^2$ and $\sigma_3^2$ are respective shadowing variance between players. $2\sigma_{2,3}$ is the correlation in shadowing which is related with $d_{2,3}$, the distance between second and third player. $X_c$ is the decorrelation distance. We assume that shadowing variance is constant across the setting. Assuming $d_2 < d_3$, first player is supposed to meet with second player. The probability of RSS at first player from the second player being larger than RSS from the third player is then:
\begin{equation}
\mathbb {P}(S_{1}^2(t) > S_{1}^3(t)) = Q\left(\dfrac{-10n_p\log\left(\dfrac{d_{1,3}}{d_{1,2}}\right)}{\sqrt{2}\sigma\sqrt{(1-e^{d_{2,3}/X_c)}}}\right)
\end{equation}
which can be upper bounded as:
\begin{equation}
\mathbb {P}(S_{1}^2(t) > S_{1}^3(t)) < Q\left(\dfrac{-10n_p\log\left(1+\dfrac{d_{2,3}}{d_{1,2}}\right)}{\sqrt{2}\sigma\sqrt{(1-e^{d_{2,3}/X_c)}}}\right)
\end{equation}
since $d_{2,3} \geq d_{1,3}-d_{1,2}$. It can be seen that it is easier for the receiver to distinguish the closer transceiver between two players when the transceivers are far from each other. When they are close to each other, the receiver does not need to separate them from each other since following any of them results in a desired behavior. Moreover, since players are taking small steps every turn, the received signals can be averaged for a finite number of last turns. Assuming players position change relatively small, the probability of players finding the closest player after N turn is:

\begin{equation}
\mathbb {P}(S_{1}^2(t)' > S_{1}^3(t)') \approx Q\left(\dfrac{-10\sqrt{N}n_p\log\left(1+\dfrac{d_{2,3}}{d_{1,2}}\right)}{\sqrt{2}\sigma\sqrt{(1-e^{d_{2,3}/X_c)}}}\right)
\end{equation}

where $S_{i}^j(t)'$ is average received RSS at player $i$ from player $j$ after $N$ turns at turn $t$. For M number of players the probability that the first player finding player $i$ as the closest player is:

\begin{equation}
\mathbb {P}_{Closest}(i) \approx \prod_{k = 2, k \neq i}^M \mathbb {P}(S_{1}^i(t)' > S_{1}^k(t)')
\end{equation}

As it can be seen, without averaging the probability of finding the optimal player decays with respect to number of players. However, for algorithms to converge, the player does not need to find the closest player. Even in the case where it tries to find a sub-optimal player, as the player gets closer to the target, the player he is trying to find becomes the optimal player. Moreover, roughly speaking the probability is higher for the players who are close to the closest player. Thus, extension to multiplayer is relatively straightforward and does not incur severe costs in terms of meeting time of the players which is also validated by simulations. 

\section{Simulation Results}
\label{simres}

\begin{figure}[!t]
  \centering
    \includegraphics[width=0.45\textwidth]{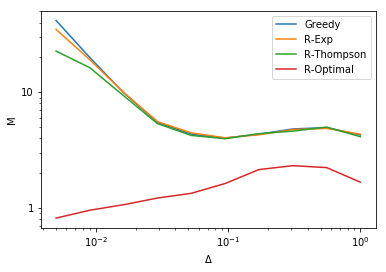}    
    \caption{Metric M vs $\Delta$ for $\sigma = 3$, $n_p = 5$, $X_c = 75$. The players are randomly initialized in a 200x200 meters area and the simulation is averaged over 1000 trials. As $\Delta$ decreases M for R-Optimal goes down and it is lower than 1 as the upper bound suggested when it is below 0.01. M for other algorithms increases when $\Delta$ is smaller than 0.1 which is also expected because of the correlation between the signals in adjacent turns.} 
    \label{fig:mdelta}
\end{figure}

The algorithms are validated with simulations. For the first set of simulations we only consider 2 players. We use the following equation as our metric for performance evaluation:

\begin{equation}
M = \dfrac{T \cdot 2\Delta - (x_t+y_t)}{x_t+y_t}
\end{equation}

where $M$ is our metric, $T$ is the averaged number of turns required for meeting, $(x_t+y_t)$ is the minimum traversed distance for players to meet. Since 2 players move $2\Delta$ every turn in total, this ratio shows the performance of the algorithm. For instance the R-Optimal upper bound goes to $2d$ as $\Delta \rightarrow 0$ as discussed earlier. Since $x_t + y_t > d$, for $d=\sqrt{x_t^2 + y_t^2}$, as $\Delta \rightarrow 0$, M is upper bounded by 1. As it can be seen from Figure \ref{fig:mdelta}, R-Optimal gets better with smaller $\Delta$ values and its $M$ value goes down under 1. As it is discussed, with small $\Delta$ values, other algorithms does not work since the information in adjacent turns get lower because of the correlation in signals. From the simulations it can be seen that the optimal value for the $\Delta$ is 0.1 for all of the algorithms but R-Optimal. In order to compare the meeting time vs number of players in the next set of simulations $\Delta$ is fixed to 0.1.

\begin{figure}[!t]
  \centering
    \includegraphics[width=0.45\textwidth]{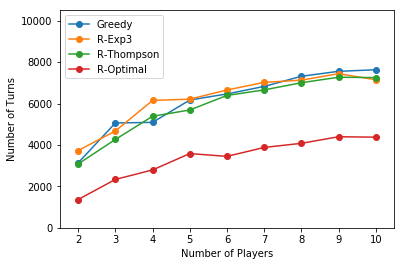}    
    \caption{Meeting time comparison of different algorithms for $\Delta = 0.1$ meters, $\sigma = 4$, $n_p = 4$ and $X_c = 75$ meters. Learning rate of 1e-3 is used for both R-Exp and R-Thompson. Randomly initialized in an area of 200x200 meters.} 
    \label{fig:4_4}
\end{figure}

\begin{figure}[!t]
  \centering
    \includegraphics[width=0.45\textwidth]{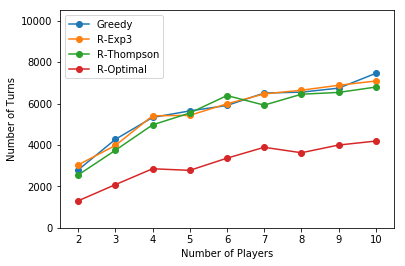}    
    \caption{Meeting time comparison of different algorithms for $\Delta = 0.1$ meters, $\sigma = 3$, $n_p = 5$ and $X_c = 75$ meters. Learning rate of 1e-3 is used for both R-Exp and R-Thompson. Randomly initialized in an area 200x200 meters.} 
    \label{fig:5_3}
\end{figure}

\begin{figure}[!t]
  \centering
    \includegraphics[width=0.45\textwidth]{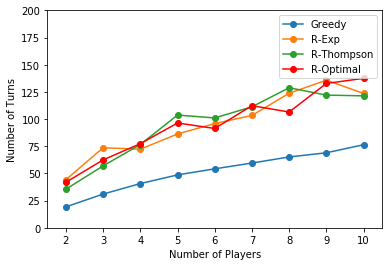}    
    \caption{Probability of finding the correct arm vs number of plays for $\Delta = 0.1$ meters, $\sigma = 3$, $n_p = 5$ and $X_c = 75$ meters. Randomly initialized in an area 2x2 meters.} 
    \label{fig:short}
\end{figure}

For the second set of simulations an area of 200x200 meters is generated and players are randomly initialized. The results are averaged for 100 trials. As it can be seen from Figure \ref{fig:4_4}, Greedy algorithm and traditional bandit algorithms have similar performances. R-Optimal algorithm gives better results as expected considering the large size of the area. In the third simulation which can be seen at Figure \ref{fig:5_3}, the shadowing variance is decreased and the path loss exponent is increased. As a result, the meeting time decreased, which is also expected from the meeting time estimations. For the last simulation a smaller area of 2x2 meters is used. As it can be seen from Figure \ref{fig:short}, Greedy algorithm works better than the other algorithms with closer players which is also expected.

\newpage

\section{Conclusion}
\label{concon}

In this paper we consider a meeting problem in mobile ad hoc network in a multi armed bandit setting where players choose one of the four possible directions to meet with each other only using RSS. We then introduce a baseline algorithm called Greedy algorithm and showed that it is comparable to algorithms which are built on top of adversarial bandit algorithms. Next, we introduce an asymptotically optimal algorithm called R-Optimal algorithm which takes advantage of the dependency between the arms and small change in signal characteristics in short periods of time due to the nature of the problem. We also give upper bounds for the expected meeting times and validate our results with simulations. 

An important observation is that the common adversarial bandit algorithms fail to solve the meeting problem. They barely decrease the meeting time compared to the Greedy algorithm. Because in the classical bandit problem, players try to minimize their regret or maximize their rewards which are observed from the arms. However in our setting, reward is observed through RSS change between turns and the distributions are non-stochastic. Under these observations it is fair to assume that the adversarial bandit algorithms can be used for our problem. Unfortunately, the distributions change more as the players get close to each other and meeting of the players becomes unlikely with traditional algorithms. The reason we extend traditional algorithms with the Greedy algorithm is to solve this problem. As future work adaptive learning rate can also be considered to solve this problem. Traditionally learning rate is decreased with respect to time. However in our case learning rate should depend on the distance between the players. Therefore, in order to have an adaptive learning rate, further assumptions might be necessary such as the knowledge of the initial power, shadowing coefficient, etc. Also, it might be impractical to assume that the players can change direction at each turn. Therefore as a future work switching cost can be added to the setting or regulated with some conditions.

\newpage

\appendix

\section{Proofs}
\subsection{Proof of Lemma 1}
\begin{proof}

The received rewards at the next turn based on RSS differences between adjacent turns can be shown with sixteen cases, based on moves of two players. These moves result in nine different probability distributions. The relation between these nine reward distributions and sixteen cases of moves are given later. Let us define $\tilde{\sigma}^2 = \sigma^2(1-e^{-2\Delta/X_c})$. The distributions are caused by the change in distance between players: 

\begin{enumerate}[label=\textbf{D.\arabic*},ref=D.\arabic*,labelindent=20pt,itemindent=1em,leftmargin=!]
\item \label{c1} $x_t+\Delta,y_t+\Delta \rightarrow \\ \mathcal{N}(-5 n_p \log\left(\dfrac{(x_{t}+\Delta)^2+(y_{t}+\Delta)^2}{x_t^2+y_t^2}\right),2\tilde{\sigma}^2) \\ = \mathcal{N}(-5 n_p \log\left(1+\dfrac{2x_t\Delta+2y_t\Delta+2\Delta^2}{x_t^2+y_t^2}\right),2\tilde{\sigma}^2)
$\\

$\approx \mathcal{N}(-5 n_p \log\left(1+\dfrac{2x_t\Delta+2y_t\Delta}{x_t^2+y_t^2}\right),2\tilde{\sigma}^2)
$ \\
Since $\sqrt{x_t^2+y_t^2} \gg \ 2\Delta$ we have $x_t^2+y_t^2 \gg 4\Delta^2$. Note that typically $\Delta < 1$. For the rest of the cases, the final approximations are given without the intermediate steps.
\item \label{c2} $x_t-\Delta,y_t+\Delta \rightarrow \\ \approx \mathcal{N}(-5 n_p \log\left(1+\dfrac{-2x_t\Delta+2y_t\Delta}{x_t^2+y_t^2}\right),2\tilde{\sigma}^2)$ 
\item \label{c3} $x_t+\Delta,y_t-\Delta \rightarrow \\ \approx \mathcal{N}(-5 n_p \log\left(1+\dfrac{2x_t\Delta-2y_t\Delta}{x_t^2+y_t^2}\right),2\tilde{\sigma}^2)$
\item \label{c4} $x_t-\Delta,y_t-\Delta \rightarrow \\ \approx \mathcal{N}(-5 n_p \log\left(1+\dfrac{-2x_t\Delta-2y_t\Delta}{x_t^2+y_t^2}\right),2\tilde{\sigma}^2)$
\item \label{c5} $x_t+2\Delta,y_i \rightarrow \\ \approx \mathcal{N}(-5 n_p \log\left(1+\dfrac{4x_t\Delta}{x_t^2+y_t^2}\right),2\tilde{\sigma}^2)$
\item \label{c6} $x_t-2\Delta,y_i \rightarrow \\ \approx \mathcal{N}(-5 n_p \log\left(1+\dfrac{-4x_t\Delta}{x_t^2+y_t^2}\right),2\tilde{\sigma}^2)$
\item \label{c7} $x_t,y_t+2\Delta \rightarrow \\ \approx \mathcal{N}(-5 n_p \log\left(1+\dfrac{4y_t\Delta}{x_t^2+y_t^2}\right),2\tilde{\sigma}^2)$
\item \label{c8} $x_t,y_t-2\Delta \rightarrow \\ \approx \mathcal{N}(-5 n_p \log\left(1+\dfrac{-4y_t\Delta}{x_t^2+y_t^2}\right),2\tilde{\sigma}^2)$
\item \label{c9} $x_t,y_t \rightarrow \mathcal{N}(0,2\tilde{\sigma}^2)$
\end{enumerate}

For the rest of the proof we can assume that $x_t,y_t \geq 0$, since the axes can be assigned in such a way that the relative distance of the second player compared to first can always be non-negative. Note that players are not aware of the axes when they play. In \ref{c1}, the relative distance between the players increase. This can be achieved with 2 different set of moves: First player moving at the -x axis with second player moving at the +y axis and first player moving at the -y axis with second player moving at the +x axis. Both of these moves result in the same distribution. The following table gives the relation between the moves and the distributions:
\begin {table}[H]
    \begin{center}
      \begin{tabular}{ | c || c | c | c | c |}
      
        \hline
        \diaghead{FirstPlayerSecondPlayer}{First Player}{Second Player} 
        & +x & -x & +y & -y \\ \hhline{=#====}
       +x & \ref{c9} & \ref{c6} & \ref{c2} & \ref{c4} \\ \hline
       -x & \ref{c5} & \ref{c9} & \ref{c1} & \ref{c3} \\ \hline
       +y & \ref{c3} & \ref{c4} & \ref{c9} & \ref{c8} \\ \hline
       -y & \ref{c1} & \ref{c2} & \ref{c7} & \ref{c9} \\
        \hline
      \end{tabular}
    \end{center}
    \caption{Moves of two players and their reward distributions}
\end{table}

The four arms are referenced as +x, -x, +y, -y for the rest of the proof. Let us define $\mathcal{I}_{i,t}$ as the chosen arm by the $i$th player at $t$th turn. We separate moves of a player in to positive and negative moves. Positive moves consist of +x and +y. Negative moves consist of -x and -y. Since $x_t,y_t \geq 0$, positive moves are the moves that the players are trying to learn. Note that a positive move of a player does not necessarily decrease the distance between players at the end of the turn since the other player may make a negative move. With the assumption of players almost moving randomly, the expected reward of a positive move in x-direction can be found as: 
\begin{equation}
\begin{split}
&\mathbb{E}[R(t+1)|\mathcal{I}_{1,t+1} = +x]  = \dfrac{-5}{4}n_p\log\left[\left(1+\dfrac{-2\Delta(x_t-y_t)}{x_t^2+y_t^2}\right)\right. \\ &\left. \left(1+\dfrac{-2\Delta(x_t+y_t)}{x_t^2+y_t^2}\right)\left(1+\dfrac{-4x_t\Delta}{x_t^2+y_t^2}\right)\right]\\& = \dfrac{-5}{4}n_p\log\left[\left(1+\dfrac{-4x_t\Delta}{x_t^2+y_t^2}+\dfrac{4\Delta^2(x_t^2-y_t^2)}{(x_t^2+y_t^2)^2}\right)\right.\\&\left.\left(1+\dfrac{-4x_t\Delta}{x_t^2+y_t^2}\right)\right] \\& \approx \dfrac{-5}{4}n_p\log\left[\left(1+\dfrac{-4x_t\Delta}{x_t^2+y_t^2}\right)\left(1+\dfrac{-4x_t\Delta}{x_t^2+y_t^2}\right)\right] \\ &= \dfrac{-10}{4}n_p\log\left(1+\dfrac{-4x_t\Delta}{x_t^2+y_t^2}\right)
\end{split}
\end{equation}
    
Since $\dfrac{4\Delta^2}{x_t^2+y_t^2} \approx 0$ and $\dfrac{x_t^2-y_t^2}{x_t^2+y_t^2} <1$, \\$\dfrac{4\Delta^2(x_t^2-y_t^2)}{(x_t^2+y_t^2)^2} = \dfrac{4\Delta^2}{x_t^2+y_t^2}\dfrac{x_t^2-y_t^2}{x_t^2+y_t^2} \approx 0$
\\

Considering the probability distribution of the reward is a summation of four Gaussian distributions,

\begin{equation}
\begin{split}
\mathbb{P}[R(t+1)|&\mathcal{I}_{1,t+1} = +x] \\ &= \mathcal{N}(-2.5n_p\log\left(1+\dfrac{-4x_t\Delta}{x_t^2+y_t^2}\right),8\tilde{\sigma}^2)
\label{eq:rdist}
\end{split}
\end{equation}

The other reward distributions are,
\begin{equation}
\begin{split}
\mathbb{P}[R(t+1)|&\mathcal{I}_{1,t+1} = -x] \\ &= \mathcal{N}(-2.5n_p\log\left(1+\dfrac{4x_t\Delta}{x_t^2+y_t^2}\right),8\tilde{\sigma}^2)
\\
\mathbb{P}[R(t+1)|&\mathcal{I}_{1,t+1} = +y] \\ &= \mathcal{N}(-2.5n_p\log\left(1+\dfrac{-4y_t\Delta}{x_t^2+y_t^2}\right),8\tilde{\sigma}^2)
\\
\mathbb{P}[R(t+1)|&\mathcal{I}_{1,t+1} = -y] \\ &= \mathcal{N}(-2.5n_p\log\left(1+\dfrac{4y_t\Delta}{x_t^2+y_t^2}\right),8\tilde{\sigma}^2)
\end{split}
    \label{eq:rdist2}
\end{equation}

Note that the means of these distributions are close to zero. Thus, the information that one can gain in a turn is relatively low, which forces players to move almost randomly. 
\end{proof}

\subsection{Proof of Lemma 2}

\begin{proof}
To estimate how many turns one would need to find the correct arms, the probability that positive arms returning positive rewards can be estimated as follows:

\begin{figure}[!t]
  \centering
    \includegraphics[width=0.45\textwidth]{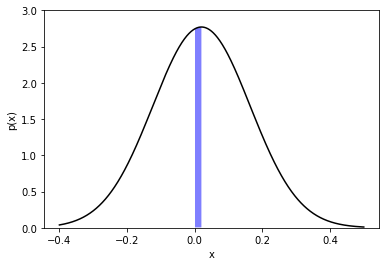}    
    \caption{Reward distribution of +x zoomed around 0  for $x_t = 20$, $d_t=30$, $\Delta = 0.1$, $\sigma = 3$ and $n_p = 8$. The shaded area is the reason the positive and negative arms can be distinguished.} 
    \label{fig:dashed_area}
\end{figure}

The shaded area at the Figure~\ref{fig:dashed_area} is the difference that allows us to differentiate the reward distributions of the arms. We are interested in the probability that a positive arm returning a positive reward which can be shown as:

\begin{equation}
\begin{split}
&\mathbb{E}[R(t+1)|\mathcal{I}_{1,t+1} = +x] \\ &= \mu_{\mathcal{I}_{1,t+1} = +x} = -2.5n_p\log\left(1+\dfrac{-4x_t\Delta}{x_t^2+y_t^2}\right) \\ &\implies 
\mathbb{P}[(R(t+1)>0)|\mathcal{I}_{1,t+1} = +x] = Q\left(\dfrac{-\mu_{\mathcal{I}_{1,t+1} = +x}}{2\sqrt{2}\tilde{\sigma}}\right) \\
&= Q\left(\dfrac{5n_p\log\left(1+\dfrac{-4x_t\Delta}{x_t^2+y_t^2}\right)}{4\sqrt{2}\tilde{\sigma}}\right) \\ &= Q\left(\dfrac{5n_p\log\left(1+\dfrac{-4x_t\Delta}{x_t^2+y_t^2}\right)}{4\sqrt{2}\sigma\sqrt{(1-e^{-2\Delta/X_c})}}\right)
\end{split}
\label{eq:trapezoid}
\end{equation}

 where $\mu_{\mathcal{I}_{i,t+1}}$ is the expected reward of the $i$th player at turn $t+1$. Let us assume that players are moving randomly but the relative position of them with respect to each other is constant. This assumption can be relaxed by forcing them to go to the initial position after finite number of turns which is always true for 2-dimensional random walks. Note that relative position changes slowly over turns which allows the relaxed assumption. With this assumption we can average the RSS changes of N turns. The averaged distribution is a Gaussian distribution with the same mean and $\dfrac{8\sigma^2}{N}$ variance. Thus the probability of averaged +x rewards being positive is: 
 
 \begin{equation}
\mathbb{P}(R(t+1)>0) =  Q\left(\dfrac{5\sqrt{N}n_p\log\left(1+\dfrac{-4x_t\Delta}{x_t^2+y_t^2}\right)}{4\sqrt{2}\sigma\sqrt{(1-e^{-2\Delta/X_c})}}\right)
\label{eq:qfunction}
 \end{equation}

Q function can be approximated by its Taylor series as:
\begin{equation}
\begin{split}
Q(x) &= \dfrac{1}{2} - \dfrac{1}{2}erf(\dfrac{x}{\sqrt{2}})\\
&= \dfrac{1}{2} - \dfrac{1}{\sqrt{\pi}}(x-\dfrac{x^3}{3}+\dfrac{x^5}{10}+...)
\end{split}
\end{equation}

In our case typically:
\begin{equation}
4\sqrt{2}\sigma\sqrt{(1-e^{-2\Delta/X_c})} \gg 5n_p\log\left(1+\dfrac{-4x_t\Delta}{x_t^2+y_t^2}\right) 
\end{equation}
Thus Q function can be approximated by a first order Taylor series:
\begin{equation}
\mathbb{P}(R(t+1)>0) \approx \dfrac{1}{2} - \dfrac{5\sqrt{N}n_p\log\left(1+\dfrac{-4x_t\Delta}{x_t^2+y_t^2}\right)}{4\sqrt{2\pi}\sigma\sqrt{(1-e^{-2\Delta/X_c})}}
\label{eq:qapprox}
\end{equation}
\end{proof}
\subsection{Proof of Theorem 3}
\begin{proof}
The expected meeting time of the greedy algorithm can be found as follows. Let us assume that $x_t$ and $y_t$ is the relative coordinates of the second player with respect to the first. Assuming an arm in x-axis is chosen, from Equation \ref{eq:rdist2} the expected rewards for +x and -x arms are:

\begin{equation}
\begin{split}
\mu_{\mathcal{I}_{1,t+1} = +x} = -2.5n_p\log\left(1+\dfrac{-4x_t\Delta}{x_t^2+y_t^2}\right) \\
\mu_{\mathcal{I}_{1,t+1} = -x} = -2.5n_p\log\left(1+\dfrac{4x_t\Delta}{x_t^2+y_t^2}\right)
\end{split}
\end{equation}

Since $\log(1+x)\approx x$ for $x \approx 0$. The rewards can be approximated as:
\begin{equation}
\begin{split}
\mu_{\mathcal{I}_{1,t+1} = +x} \approx \dfrac{10 n_p x_t\Delta}{x_t^2+y_t^2} \\
\mu_{\mathcal{I}_{1,t+1} = -x} \approx \dfrac{-10 n_p x_t\Delta}{x_t^2+y_t^2}
\end{split}
\label{eq:rew}
\end{equation}

Note that $\mu_{+x}$ is small initially and we are not averaging signals as in the Lemma 2. Therefore, the probability that a positive arm having a positive reward can be approximated as Equation \ref{eq:qapprox} where $N = 1$ and where we replace the logarithm with the approximation. A similar equation can be used for the negative arm. The resulting equations are:

\begin{equation}
\begin{split}
\mathbb{P}[(R(t+1)>0)|&\mathcal{I}_{1,t+1} = +x] \\
&\approx \dfrac{1}{2} + \dfrac{5n_px_t\Delta}{\sqrt{2\pi}(x_t^2+y_t^2)\sigma\sqrt{(1-e^{-2\Delta/X_c})}}\\
\mathbb{P}[(R(t+1)>0)|&\mathcal{I}_{1,t+1} = -x] \\
&\approx \dfrac{1}{2} - \dfrac{5n_px_t\Delta}{\sqrt{2\pi}(x_t^2+y_t^2)\sigma\sqrt{(1-e^{-2\Delta/X_c})}}
\end{split}
\end{equation}

Note that the above estimation is different than the reward averaging where the variance decrease with N. The Greedy algorithm only depends on the last turn of the players. Thus, it can be represented as a Markov chain with the state probabilities using the estimated rewards:
\FloatBarrier
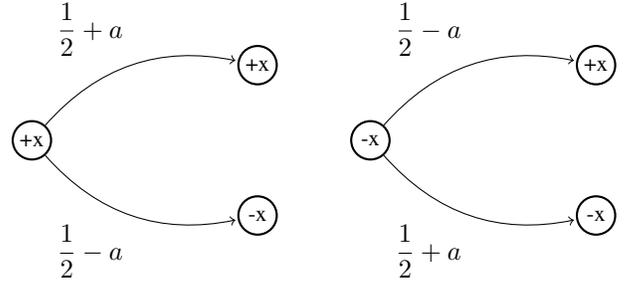
\begin{figure}[!h]
\begin{center}
\begin{tikzpicture}[scale=1,shorten >=1pt, auto, node distance=1cm,
   node_style/.style={scale=0.85,circle,draw=black,thick,inner sep=0pt,minimum size=0.6cm}]
   % Drawing goes here
   \node[node_style] at (-3, 0) (1st) {+x};
   \node[node_style] at (0, 1) (2nd+) {+x};
   \node[node_style] at (0, -1) (2nd-) {-x};
   \node[node_style] at (1.5, 0) (1st2) {-x};
   \node[node_style] at (4.5, 1) (2nd+2) {+x};
   \node[node_style] at (4.5, -1) (2nd-2) {-x};
    \draw[every loop]
            (1st) edge[bend left, auto] node {$\dfrac{1}{2}+a$} (2nd+)
            (1st2) edge[bend left, auto = left] node {$\dfrac{1}{2}-a$} (2nd+2)
            (1st) edge[bend right, auto = right] node {$\dfrac{1}{2}-a$} (2nd-)
            (1st2) edge[bend right, auto = right] node {$\dfrac{1}{2}+a$} (2nd-2);
\end{tikzpicture}
\end{center}
\caption{Markov chain representation of Greedy algorithm cycle, the initial states are the random states and the other states are decided based on the sign of the rewards. $a=\dfrac{5n_px_t\Delta}{\sqrt{2\pi}(x_t^2+y_t^2)\sigma\sqrt{(1-e^{-2\Delta/X_c})}}$.}
\label{fig:markov1}
\end{figure}

As it can be seen from Figure ~\ref{fig:markov1}, assuming an arm in x direction is chosen, after two turns the probability of returning to the initial position is:
\begin{equation}
\label{eq:no_change}
P_{x_t,y_t} = \dfrac{1}{2}(\dfrac{1}{2}-a) + \dfrac{1}{2}(\dfrac{1}{2}+a) = \dfrac{1}{2}
\end{equation}

where $a$ is $\dfrac{5n_px_t\Delta}{\sqrt{2\pi}(x_t^2+y_t^2)\sigma\sqrt{(1-e^{-2\Delta/X_c})}}$. First term at the right hand side is the probability of choosing the +x arm and receiving a negative reward. Second term is the probability of choosing the -x arm and receiving a negative reward. With same approach, after two turns the probability of having ($x_t-2\Delta,y_t$) and ($x_t+2\Delta,y_t$) relative positions are:

\begin{equation}
\label{eq:change}
\begin{split}
P_{x_t+2\Delta,y_t} &= \dfrac{1}{2}(\dfrac{1}{2}-a) \\ &= \dfrac{1}{4}-\dfrac{5n_px_t\Delta}{2\sqrt{2\pi}(x_t^2+y_t^2)\sigma\sqrt{(1-e^{-2\Delta/X_c})}}\\
P_{x_t-2\Delta,y_t} &= \dfrac{1}{2}(\dfrac{1}{2}+a) \\ &= \dfrac{1}{4}+\dfrac{5n_px_t\Delta}{2\sqrt{2\pi}(x_t^2+y_t^2)\sigma\sqrt{(1-e^{-2\Delta/X_c})}}
\end{split}
\end{equation}

Players will move towards each other since $P_{x_t-2\Delta,y_t} > P_{x_t+2\Delta,y_t}$ given that there is enough number of turns. The second term at the right hand size, $\dfrac{5n_px_t\Delta}{2\sqrt{2\pi}(x_t^2+y_t^2)\sigma\sqrt{(1-e^{-2\Delta/X_c})}}$, increases gradually as the players move towards each other. Thus the expected number of turns can be upper bounded assuming it as a constant. Assume that N is the total number of turns, then $\dfrac{N}{2}$ is the number of greedy algorithm cycles. Therefore:

\begin{equation}
\begin{split}
&x_{t+2N} - x_t = \dfrac{N}{2}\dfrac{1}{2}(0)+\dfrac{N}{2}(\dfrac{1}{4}+a)(-2\Delta)\\&+\dfrac{N}{2}(\dfrac{1}{4}-a)(2\Delta) = -\dfrac{5Nn_px_t\Delta^2}{\sqrt{2\pi}(x_t^2+y_t^2)\sigma\sqrt{(1-e^{-2\Delta/X_c})}}
\end{split}
\end{equation}
In order for players to meet at the center with respect to each other, first player needs to have $x_{t+2N} = \dfrac{x_t}{2}$, Expected N can be upper bounded with:
\begin{equation}
N = \dfrac{(x_t^2+y_t^2)\sqrt{2\pi\sigma^2(1-e^{-2\Delta/X_c})}}{10 n_p \Delta^2}
\end{equation}

 Note that $x_{t+2N}$ and $y_{t+2N}$ need to be equal to zero at the same turn for the meeting condition. The upper bound is symmetric with respect to $x_t$ and $y_t$. Since the arms in the x-direction is pulled half of the time, the final expected number of turns for meeting can be upper bounded with $\dfrac{(x_t^2+y_t^2)\sqrt{2\pi\sigma^2(1-e^{-2\Delta/X_c})}}{5 n_p \Delta^2}$ turns. 
 
\end{proof}

\subsection{Proof of Theorem 4}
\begin{proof}
The R-Optimal algorithm uses mean of the rewards which creates a Markovian state representation of the problem. This should not be confused with traditional definition of Markovian multi armed bandit problem in which the rewards are assumed to be Markovian \cite{anantharam1987asymptotically}, \cite{tekin2010online}. However in our case reward distribution change at each turn depending on decisions of players. That is why having a Markovian reward model is not a proper approach. Instead the state that players are in can be modeled with a Markov chain to analyze the algorithm's estimated time of meeting.

In order to represent the state that the algorithm is in, we are interested in the mean rewards of the arms which are in the reverse direction. Just as in the previous proofs, we consider only the x-axis and generalize the result later. The states of the R-Optimal algorithm can be shown as follows to upper bound the expected number of turns asymptotically. Let us define $\lambda_i(t) = k_{i,+}(t)-k_{i,-}(t)$, where $k_{i,+}(t)$ and $k_{i,-}(t)$ are the mean of rewards for the positive and negative arms for player $i$ at turn $t$ in x-axis respectively.  At turn $t$, the probabilities of a positive arm receiving a positive reward and a negative arm receiving a negative reward can be used to construct the Markov chain:

\begin{equation}
\begin{split}
\mathbb{P}[(R(t+1)>0)|&\mathcal{I}_{1,t+1} = +x]\\
&\approx \dfrac{1}{2} + \dfrac{5n_px_t\Delta}{\sqrt{2\pi}(x_t^2+y_t^2)\sigma\sqrt{(1-e^{-2\Delta/X_c})}}\\
\mathbb{P}[(R(t+1)>0)|&\mathcal{I}_{1,t+1} = -x]\\
&\approx \dfrac{1}{2} - \dfrac{5n_px_t\Delta}{\sqrt{2\pi}(x_t^2+y_t^2)\sigma\sqrt{(1-e^{-2\Delta/X_c})}}
\end{split}
\end{equation}

Note that similar to the upper bound of the Greedy algorithm, the second term at the right hand size is treated as a constant for the rest of the proof to find an upper bound, since it gradually increases as players move towards each other. Second term at the right hand size is denoted with $a$ for the rest of the proof. Since there are two options for the initial chosen arm, there are 2 possible Markov chains which can be constructed as in Figure~\ref{fig:markov2}.

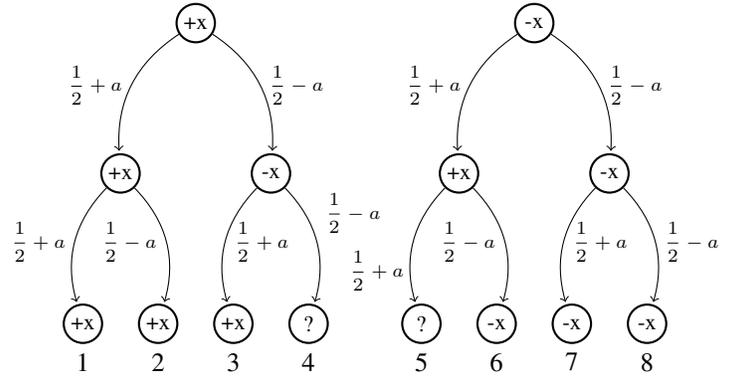
\begin{figure}[!h]
\begin{center}
\begin{tikzpicture}[scale=1,shorten >=1pt, auto, node distance=1cm,
   node_style/.style={scale=0.85,circle,draw=black,thick,inner sep=0pt,minimum size=0.6cm}]
   % Drawing goes here
   \node[node_style] at (-3, 0) (left_top) {+x};
   \node[node_style] at (-4, -2) (left_middle_1) {+x};
   \node[node_style] at (-2, -2) (left_middle_2) {-x};
   \node[node_style,label=below:1] at (-4.5, -4) (left_bot_1) {+x};
   \node[node_style,label=below:2] at (-3.5, -4) (left_bot_2) {+x};
   \node[node_style,label=below:3] at (-2.5, -4) (left_bot_3) {+x};
   \node[node_style,label=below:4] at (-1.5, -4) (left_bot_4) {?};
   \node[node_style] at (1.5, 0) (right_top) {-x};
   \node[node_style] at (2.5, -2) (right_middle_1) {-x};
   \node[node_style] at (0.5, -2) (right_middle_2) {+x};
   \node[node_style,label=below:8] at (3, -4) (right_bot_1) {-x};
   \node[node_style,label=below:7] at (2, -4) (right_bot_2) {-x};
   \node[node_style,label=below:6] at (1, -4) (right_bot_3) {-x};
   \node[node_style,label=below:5] at (0, -4) (right_bot_4) {?};
    \draw[every loop]
            (left_top) edge[bend right, left, font=\fontsize{7}{12}] node { $\dfrac{1}{2}+a$} (left_middle_1)
            (left_top) edge[bend left, right, font=\fontsize{7}{12}] node {$\dfrac{1}{2}-a$} (left_middle_2)
            (left_middle_1) edge[bend right, left, font=\fontsize{7}{12}] node {$\dfrac{1}{2}+a$} (left_bot_1)
            (left_middle_1) edge[bend left, left, font=\fontsize{7}{12}] node {$\dfrac{1}{2}-a$} (left_bot_2)
            (left_middle_2) edge[bend right, right, font=\fontsize{7}{12}] node {$\dfrac{1}{2}+a$} (left_bot_3)
            (left_middle_2) edge[bend left, auto, font=\fontsize{7}{12}] node {$\dfrac{1}{2}-a$} (left_bot_4)
            (right_top) edge[bend left, right, font=\fontsize{7}{12}] node { $\dfrac{1}{2}-a$} (right_middle_1)
            (right_top) edge[bend right, left, font=\fontsize{7}{12}] node {$\dfrac{1}{2}+a$} (right_middle_2)
            (right_middle_1) edge[bend left, right, font=\fontsize{7}{12}] node {$\dfrac{1}{2}-a$} (right_bot_1)
            (right_middle_1) edge[bend right, right, font=\fontsize{7}{12}] node {$\dfrac{1}{2}+a$} (right_bot_2)
            (right_middle_2) edge[bend left, left, font=\fontsize{7}{12}] node {$\dfrac{1}{2}-a$} (right_bot_3)
            (right_middle_2) edge[bend right, below left, font=\fontsize{7}{12}] node {$\dfrac{1}{2}+a$} (right_bot_4);
\end{tikzpicture}
\end{center}
\caption{Markov chain representation of R-Optimal algorithm cycle, the initial states are the maximum mean reward states, the second state is decided based on sign of the rewards and third state is again the maximum mean reward state. $a=\dfrac{5n_px_t\Delta}{\sqrt{2\pi}(x_t^2+y_t^2)\sigma\sqrt{(1-e^{-2\Delta/X_c})}}$. States which are denoted with "?" depend on the initial states(top states). In other words, those states depend on the $\lambda_i(t)$ which is the difference between the means of reverse arms for the initial states.}
\label{fig:markov2}
\end{figure}

Let us define $p_{i,+}(t)$, $p_{i,-}(t)$, $c_{i,+}(t)$, $c_{i,-}(t)$ where $p_{i,+}(t)$ and $p_{i,-}(t)$ are the total number of positive rewards; $c_{i,+}(t)$ and $c_{i,-}(t)$ are the total number of plays for positive and negative arms for player $i$ at turn $t$ in x-axis respectively. After pulling +x twice, with probability $(\dfrac{1}{2}+a)^2$, the player receives two positive rewards which corresponds to the path reaching state 1 in Figure~\ref{fig:markov2}. Note that asymptotically, $2p_{i,+}(t) \approx c_{i,+}(t)$ and $2p_{i,-}(t) \approx c_{i,-}(t)$, since rewards' means are close to zero. Thus,
\begin{equation}
\begin{split}
\lambda_1(t+2) &= \dfrac{p_{1,+}(t)+2}{c_{1,+}(t)+2} - \dfrac{p_{1,-}(t)}{c_{1,-}(t)} \\ &= \lambda_1(t) + \dfrac{p_{1,+}(t)+2}{c_{1,+}(t)+2} - \dfrac{p_{1,+}(t)}{c_{1,+}(t)}\\
&=\lambda_1(t) + \dfrac{2(c_{1,+}(t)-p_{1,+}(t))}{c_{1,+}(t)(c_{1,+}(t)+2)} \approx \lambda_1(t) + \dfrac{1}{c_{1,+}(t)}\\ &= \lambda_1(t) + \delta_+
\end{split}
\end{equation}
for $\delta_+ = \dfrac{1}{c_{1,+}(t)}$ and $c_{1,+}(t) \gg 2$. Complementary case for the path which reaches state 8 has 2 consecutive -x pulls. It has $\lambda_1(t+2)$ = $\lambda_1(t) - \delta_-(t)$ for $\delta_- = \dfrac{1}{c_{1,-}}$ and $c_{1,-}(t) \gg 2$ with probability $(\dfrac{1}{2}-a)^2$. For the path which reaches to state 2 in Figure~\ref{fig:markov2}, with probability $\dfrac{1}{4}-a^2$:

\begin{equation}
\begin{split}
\lambda_1(t+2) &= \dfrac{p_{1,+}(t)+1}{c_{1,+}(t)+2} - \dfrac{p_{1,-}(t)}{c_{1,-}(t)} \\ &= \lambda_1(t) + \dfrac{p_{1,+}(t)+1}{c_{1,+}(t)+2} - \dfrac{p_{1,+}(t)}{c_{1,+}(t)}\\
&=\lambda_1(t) + \dfrac{c_{1,+}(t)-2p_{1,+}(t)}{c_{1,+}(t)(c_{1,+}(t)+2)} \approx \lambda_1(t) 
\end{split}
\end{equation}

Similar equations give the same result for the paths which reach to states 3,6,7 all with probability $\dfrac{1}{4}-a^2$. For the path which ends at state 4:

\begin{equation}
\begin{split}
\lambda_1(t+2) &= \dfrac{p_{1,+}(t)}{c_{1,+}(t)+1} - \dfrac{p_{1,-}(t)+1}{c_{1,-}(t)+1}\\ 
&= \lambda_1(t) + \dfrac{p_{1,+}(t)}{c_{1,+}(t)+1} - \dfrac{p_{1,+}(t)}{c_{1,+}(t)} \\ &- \dfrac{p_{1,-}(t)+1}{c_{1,-}(t)+1} + \dfrac{p_{1,-}(t)}{c_{1,-}(t)}\\
&\approx \lambda_1(t) - \dfrac{\delta_+}{2} - \dfrac{\delta_-}{2} \approx \lambda_1(t) - \delta
\end{split}
\end{equation}

It can be assumed for all of the states that $\delta = \delta_- \approx \delta_+$, since we are trying to find when the algorithm consistently picks one arm more than the other. Until that point, it can be assumed that the arms will be picked roughly similar amount of times. Even when players find each other, the number of pulls of positive and negative arms are similar to each other considering high amount of turns are required for the problem. That is why it results in similar number of pulls. 

With a similar calculation, state 5 has $\lambda_1(t+2) \approx \lambda_1(t) + \delta$ with probability $\dfrac{1}{4}+a^2$. Pulled arm at state 4 and 5 depend on the magnitude of $\lambda_1(t)$, since algorithm chooses an arm for state 4 and 5 depending on the sign of $\lambda_1(t) - \delta$ and $\lambda_1(t) + \delta$ respectively. For other cases the sign of $\lambda_1(t+2)$ is equal to the sign of $\lambda_1(t)$. Using these equations the Markov chain can be simplified to:

\begin{figure}[!h]
\begin{center}
\begin{tikzpicture}[scale=1,shorten >=1pt, auto, node distance=1cm,
   node_style/.style={scale=0.85,circle,draw=black,thick,inner sep=0pt,minimum size=0.6cm}]
   % Drawing goes here
   \node[node_style,label=below:$\lambda_1(t)$] at (-3, 0) (left_top) {+x};
   \node[node_style,label=below:$\lambda_1(t)+\delta$] at (-4, -2) (left_middle_1) {+x};
   \node[node_style,label=below:$\lambda_1(t)-\delta$] at (-2, -2) (left_middle_2) {?};
   
   \node[node_style,label=below:$\lambda_1(t)$] at (1.5, 0) (right_top) {-x};
   \node[node_style,label=below:$\lambda_1(t)-\delta$] at (2.5, -2) (right_middle_1) {-x};
   \node[node_style,label=below:$\lambda_1(t)+\delta$] at (0.5, -2) (right_middle_2) {?};

    \draw[every loop]
            (left_top) edge[bend right, left, font=\fontsize{7}{12}] node { $(\dfrac{1}{2}+a)^2$} (left_middle_1)
            (left_top) edge[bend left, right, font=\fontsize{7}{12}] node {$(\dfrac{1}{2}-a)^2$} (left_middle_2)
            (left_top) edge[loop above, left, font=\fontsize{7}{12}] node { $\dfrac{1}{2}-2a^2$} (left_top)

            (right_top) edge[bend left, right, font=\fontsize{7}{12}] node { $(\dfrac{1}{2}-a)^2$} (right_middle_1)
            (right_top) edge[bend right, left, font=\fontsize{7}{12}] node {$(\dfrac{1}{2}+a)^2$} (right_middle_2)
            (right_top) edge[loop above, left, font=\fontsize{7}{12}] node { $\dfrac{1}{2}-2a^2$} (right_top);

\end{tikzpicture}
\end{center}
\caption{Markov chain representation of R-Optimal algorithm cycle where each state represents a cycle. $a = \dfrac{\mu_{\mathcal{I}_{1,t+1} = +x}}{{4\sqrt{\pi\sigma^2}}}$}
\label{fig:markov3}
\end{figure}

Note that, while returning to the same state with probability $\dfrac{1}{2}-2a^2$, there is $\dfrac{1}{2}-a$ chance of pulling the negative arm and $\dfrac{1}{2}+a$ chance of pulling the positive arm. Thus, asymptotically self-returning moves also help for the meeting of the nodes. However, the effect is relatively small and for the upper bound it can be ignored. Combining all of the aforementioned observations, the Markov chain in Figure~\ref{fig:markov3} can be constructed.

\begin{figure}[!h]
\begin{center}

\begin{tikzpicture}[scale=0.95,shorten >=1pt, auto, node distance=1cm,
   node_style/.style={scale=0.85,circle,draw=black,thick,inner sep=0pt,minimum size=0.5cm}]
   % Drawing goes here
   \node[node_style,label=below:{\scriptsize $\mathbf{3\pmb{\delta}}$}] at (-4.2, -2) (left_3) {+x};
      \node[node_style,label=below:{\scriptsize$\mathbf{2\pmb{\delta}}$}] at (-2.8, -2) (left_2) {+x};
   \node[node_style,label=below:{\scriptsize$\mathbf{\pmb{\delta}}$}] at (-1.4, -2) (left_1) {+x};
   \node[node_style,label=below:{\scriptsize$\mathbf{0}$}] at (0, -2) (middle) {+x};
    \node[node_style,label=below:{\scriptsize$\mathbf{-\pmb{\delta}}$}] at (1.4, -2) (right_1) {-x};
    \node[node_style,label=below:{\scriptsize$\mathbf{-2\pmb{\delta}}$}] at (2.8, -2) (right_2) {-x};
   \node[node_style,label=below:{\scriptsize$\mathbf{-3\pmb{\delta}}$}] at (4.2, -2) (right_3) {-x};

    \draw[every loop]
            (middle) edge[bend left, below, font=\fontsize{5}{12}] node { $(\dfrac{1}{2}+a)^2$} (left_1)
            (middle) edge[bend left, above, font=\fontsize{5}{12}] node {$(\dfrac{1}{2}-a)^2$} (right_1)
            (middle) edge[ out=105, in=75
                , looseness=0.8, loop
                , distance=1cm, font=\fontsize{5}{12}, ->] node { $\dfrac{1}{2}-2a^2$} (middle)

            (right_1) edge[bend left, below, font=\fontsize{5}{12}] node { $(\dfrac{1}{2}+a)^2$} (middle)
            (right_1) edge[bend left, above, font=\fontsize{5}{12}] node {$(\dfrac{1}{2}-a)^2$} (right_2)
            (right_1) edge[ out=105, in=75
                , looseness=0.8, loop
                , distance=1cm, font=\fontsize{5}{12}, ->] node { $\dfrac{1}{2}-2a^2$} (right_1)

            (right_2) edge[bend left, below, font=\fontsize{5}{12}] node { $(\dfrac{1}{2}+a)^2$} (right_1)
            (right_2) edge[bend left, above, font=\fontsize{5}{12}] node {$(\dfrac{1}{2}-a)^2$} (right_3)
            (right_2) edge[ out=105, in=75
                , looseness=0.8, loop
                , distance=1cm, font=\fontsize{5}{12}, ->]node { $\dfrac{1}{2}-2a^2$} (right_2)

            (right_3) edge[bend left, below, font=\fontsize{5}{12}] node { $(\dfrac{1}{2}+a)^2$} (right_2)
            (right_3) edge[ out=105, in=75
                , looseness=0.8, loop
                , distance=1cm, font=\fontsize{5}{12}, ->] node { $\dfrac{1}{2}-2a^2$} (right_3)
            
            (left_1) edge[bend left, above, font=\fontsize{5}{12}] node { $(\dfrac{1}{2}-a)^2$} (middle)
            (left_1) edge[bend left, below, font=\fontsize{5}{12}] node {$(\dfrac{1}{2}+a)^2$} (left_2)
            (left_1) edge[ out=105, in=75
                , looseness=0.8, loop
                , distance=1cm, font=\fontsize{5}{12}, ->] node { $\dfrac{1}{2}-2a^2$} (left_1)

            (left_2) edge[bend left, above, font=\fontsize{5}{12}] node { $(\dfrac{1}{2}-a)^2$} (left_1)
            (left_2) edge[bend left, below, font=\fontsize{5}{12}] node {$(\dfrac{1}{2}+a)^2$} (left_3)
            (left_2) edge[ out=105, in=75
                , looseness=0.8, loop
                , distance=1cm, font=\fontsize{5}{12}, ->] node { $\dfrac{1}{2}-2a^2$} (left_2)

            (left_3) edge[bend left, above, font=\fontsize{5}{12}] node { $(\dfrac{1}{2}-a)^2$} (left_2)
            (left_3) edge[ out=105, in=75
                , looseness=0.8, loop
                , distance=1cm, font=\fontsize{5}{12}, ->] node { $\dfrac{1}{2}-2a^2$} (left_3);

\end{tikzpicture}
\end{center}
\caption{Markov chain representation of R-Optimal algorithm cycle where each state represents a cycle. $a = \dfrac{\mu_{\mathcal{I}_{1,t+1} = +x}}{{4\sqrt{\pi\sigma^2}}}$}
\label{fig:markov3}
\end{figure}

To prove that this chain eventually converges to a state with +x starting from anywhere, let $P_{z,z-1}$ be the probability of reaching state $z-1$ starting from $z$. We assume that the states with +x correspond to $z = -1,-2,-3,...$. Since the chain is symmetric; we have $P_{z,z-1} = P_{z-1,z-2} = ... =  P_{0,-1} = P$. Thus reaching state $z-2$ starting from $z$ can be found as, $P_{z,z-2} = P_{z,z-1}P_{z-1,z-2} = P^2$. Hence, the following equation can be solved to find P,

\begin{equation}
P=P(\dfrac{1}{2}-2a^2)+P^2(\dfrac{1}{2}-a)^2+(\dfrac{1}{2}+a)^2
\end{equation}

The equation has two roots, $r_1 = 1$ and $r_2 = \dfrac{1+4a^2+4a}{1+4a^2-4a}$. Since $r_2 > 1$, $r_1$ is the valid root. Thus, $P=1$. In other words, starting from anywhere it is guaranteed to move one state left. This drift guarantees that eventually a state with +x will be reached with probability 1.

Feller shows that starting from position $z$ with boundary conditions $(0,b)$ where $0<z<b$, th random walk is expected to end in \cite{feller2008introduction},
\begin{equation}
T = \dfrac{z}{q-p} - \dfrac{b}{q-p}\dfrac{1-(q/p)^z}{1-(q/p)^b}
\end{equation}
where boundary condition means expected time to reach either $0$ or $b$. $p$ is the probability to move towards $b$ and $q$ is the probability to move towards $0$ with $p+q=1$. In our case there is no upper bound. Thus for $b = \infty$ and $q>p$ in the limit, $T\approx\dfrac{z}{q-p}$. Since self returning moves have no effect on convergence to the positive arms or to the estimated meeting time, it can be ignored and the estimated time can be scaled. Therefore, 
\begin{equation}
\begin{split}
p&=\dfrac{(\dfrac{1}{2}-a)^2}{\dfrac{1}{2}+2a^2}\quad q =\dfrac{(\dfrac{1}{2}+a)^2}{\dfrac{1}{2}+2a^2}\\
\implies T &= z\dfrac{4a^2+1}{4a}
\end{split}
\end{equation}
Considering self returning moves happening with probability $\dfrac{1}{2}-2a^2$, estimated time to reach state $\delta$ starting from state 0 is:
\begin{equation}
\begin{split}
T_{estimated} &= \dfrac{4a^2+1}{4a}\dfrac{2}{4a^2+1} = \dfrac{1}{2a} \\
&= \dfrac{(x_t^2+y_t^2)\sqrt{2\pi\sigma^2(1-e^{-2\Delta/X_c})}}{10n_px_t\Delta}
\end{split}
\end{equation}

Since the initial state is expected to be state 0, estimation is calculated based on it. After a state with positive arm is reached, positive arm is chosen consistently more than the negative arm. Assuming it stays at a positive state, the negative arm can only be chosen during the exploration step where the reverse arm is pulled depending on the reward. Note that, exploration step is crucial to have since we assume $\delta_+ \approx \delta_-$. Thus exploration step, where we pull the reverse arm, fastens the convergence to a positive state. However player might not stay at a positive state. After a player reaches a positive state, it can still move back to a negative state. In order to upper bound the estimated time of meeting, $z \gg 1$ should be chosen so that player stays at positive states with high probability. Intuitively, it corresponds to a player reaching a high positive $\delta$ state so it becomes unlikely to move back to a negative state. Until the player reaches to this state, the algorithm can be considered as Greedy algorithm in terms of expected meeting time.

Expected meeting time is upper bounded considering both the x-axis and y-axis. The expected time can be upper bounded with the maximum of $|x_t|$ and $|y_t|$, since for the axis which is further away is expected to take more time than the other. Then, assuming $|x_t| > |y_t|$ (the same proof can be followed for the other case):

\begin{equation}
\begin{split}
&T_{estimated} = \dfrac{(x_t^2+y_t^2)\sqrt{2\pi\sigma^2(1-e^{-2\Delta/X_c})}}{10n_px_t\Delta} \\
&< \dfrac{(2x_t^2)\sqrt{2\pi\sigma^2(1-e^{-2\Delta/X_c})}}{10n_px_t\Delta} \\
&= \dfrac{(2x_t)\sqrt{2\pi\sigma^2(1-e^{-2\Delta/X_c})}}{10n_p\Delta} < \dfrac{d\sqrt{2\pi\sigma^2(1-e^{-2\Delta/X_c})}}{5n_p\Delta}
\end{split}
\end{equation}

where $d$ is $\sqrt{x_t^2+y_t^2}$. Then the expected meeting time can be upper bounded as,
\begin{equation}
\begin{split}
N &= 2(zT_{estimated} + \dfrac{d}{\Delta}) \\ &= \dfrac{2zd\sqrt{2\pi\sigma^2(1-e^{-2\Delta/X_c})}}{5n_p\Delta} + \dfrac{d}{\Delta}
\end{split}
\end{equation}

where $z$ is chosen to keep the players at positive states. Empirically for z = 10, the upper bound holds with very high probability. The second term comes from the fact that after a positive state is maintained, the players are closer to each other than initial positions. Because, self returning moves have a higher chance of pulling a positive arm. For the upper bound, we can ignore this and assume that players do not get closer until a positive state is maintained. Therefore, in order to find the upper bound with respect to x-axis, the distance a player needs to travel is $\dfrac{x_t}{2}$ which can be upper bounded with $\dfrac{d}{2}$, considering $d \geq x_t$ and $d \geq y_t$.  Thus, expected turn can be upper bounded with $\dfrac{d}{\Delta}$. Constant 2 is caused by the fact that there are two axes.
\end{proof}

\ifCLASSOPTIONcaptionsoff
  \newpage
\fi

\newpage
\bibliographystyle{IEEEtran}  
\bibliography{references}

\end{document}